\documentstyle[12pt,preprint,aps]{revtex}

\begin{document}

\draft

\title{Is Anomalous Production of $\Omega$ and $\overline{\Omega}$
Evidence for Disoriented Chiral Condensates?}

\author{J.I. Kapusta$^1$ and S.M.H. Wong$^{1,2}$}

\address{$^1$School of Physics and Astronomy, University of Minnesota, 
Minneapolis, MN 55455 \\
$^{2,\dag}$Department of Physics, Ohio State University, Columbus, OH 43210}

\date{\today}

\maketitle

\begin{abstract}
No conventional picture of nucleus-nucleus collisions has yet been able
to explain the abundance of $\Omega$ and $\overline{\Omega}$ hadrons in
central collisions between Pb nuclei at 158 A GeV at the CERN SPS.  We
argue that this is evidence that they are produced as
topological defects arising from the formation of disoriented chiral
condensates (DCC) with an average domain size of about 2 fm. 
\end{abstract}

\pacs{PACS numbers: 25.75.Dw, 11.27.+d, 24.85.+p
\hfill NUC-MINN-2000/25-T}

One of the most interesting observations in the field of high energy
heavy ion physics is that the hadronic phase space is populated
statistically (for the most recent analyses see \cite{freeze1,freeze2}).
This means that the relative abundances of a wide variety 
of hadrons, such as protons, singly and doubly strange hyperons, 
pions, kaons, eta-mesons, and the antiparticles of these can
all be described quite well in terms of a common temperature and chemical
potentials (for electric charge, baryon number, and net strangeness).
The time at which this occurs is referred to as chemical freezeout.
For central Pb+Pb collisions at a beam energy of 158 A GeV at the CERN
SPS, for example, analyses typically give this temperature in the range
175$\pm$10 MeV.  How this population is achieved dynamically is still a
matter of much debate.  There are proponents for hadronization from an
earlier state of quark-gluon plasma, and there are proponents for
strong interactions among the hadronic degrees of freedom alone without
recourse to partonic ones \cite{qm}.  However, the abundances of the
$\Omega$ and $\overline{\Omega}$, which contain three strange quarks or
antiquarks, are an anomaly.  The purpose of this letter is to put
forward the argument that this anomaly is most easily understood if a
disoriented chiral condensate (DCC) is formed.  Formation of DCC is
expected on quite general grounds \cite{DCC1,DCC2,DCC3,DCC4}
but up to now there has been no experimental evidence for it.

The anomaly is manifest in several ways.  First, one obtains a chi-squared
fit to the particle ratios which is much better if the
$\Omega$ and $\overline{\Omega}$ are left out of the fitting procedure
\cite{RandL}.  This results in a lower chemical freezeout temperature
of 145$\pm$5 MeV and a prediction for the $\Omega$ and $\overline{\Omega}$ 
abundancies which are smaller than observations by a factor of 2.
Second, an analysis of chemical equilibration times based on experimentally
measured hadronic reactions indicates that the more strange valence
quarks a hadron contains the longer it takes to establish chemical
equilibrium \cite{sreview}.  Pions and kaons easily equilibrate.  Even
the time to equilibrate the antiproton is short enough, about 3 fm/c,
to realize in a heavy ion collision on account of the many-body
reaction $5\pi \rightarrow p+\overline{p}$ \cite{sreview,ed}.  The
equilibration times for the $\Omega$ and $\overline{\Omega}$ are
longest of all, being on the order of several hundred fm/c compared to
the canonical nucleus-nucleus collision time of 10 fm/c \cite{qm}.
This is the result of numerically solving a coupled set of rate
equations for hadrons.  The reason it is so long is that the $\Omega$
(or $\overline{\Omega}$) may be produced via a sequence of
two-body reactions like $K+N\rightarrow \Lambda+\pi$,
$K+\Lambda\rightarrow \Xi+\pi$, $K+\Xi\rightarrow \Omega+\pi$, etc.
and since the $\Omega$ is at the end of the chain it takes the
longest to build up its abundance.  Multi-body reactions like
$3K+2\pi \rightarrow \Omega +\overline{\Omega}$ help, and indeed were
included in the rate equations, but these are slow compared to
the buildup of antiprotons because the number of kaons is smaller
than pions by about a factor of 5. Third, the microscopic
transport model UrQMD \cite{urqmd}, which is generally regarded
to be the most sophisticated such model available, cannot reproduce
the high yields of hyperons, especially the $\Omega$ and
$\overline{\Omega}$.  For those it falls short by about a factor of
2 to 3 for central Pb+Pb collisions at the SPS \cite{surqmd}.
One way to account for the increased hyperon production within UrQMD
is to lower the constituent quark masses to the values of the current
quarks.  Another is to increase the string tension by a factor of 3.
The authors admit that both approaches are rather ad hoc, and go on
to say that the high yield of hyperons ``is so far the only
clear signal which contradicts transport calculations based on hadronic
and string degrees of freedom".  This may be true insofar as only two-body
initial states are treated by UrQMD whereas detailed balance would require
the inclusion of all time-reversed reactions \cite{detail}, as in the
case of the antiproton mentioned above.

We now propose that this anomaly arises
from the formation of topological defects in DCC.  These
topological defects are identified with Skyrmions \cite{skyrme}.
Such a mechanism for the production of baryons and antibaryons in
heavy ion collisions was developed in a series of papers
\cite{degrand,heinz,ajit,sb} but never applied quantitatively to
data.  The basic idea is that the chiral field (identified with
pions if two flavors and with pions and kaons if three flavors)
may become completely disoriented in coordinate space beyond a
characteristic distance $\xi$, the correlation length or domain
size.  This disorientation may occur in a heavy ion collision if the
entropy is very large, as one might easily imagine in such a violent
event.  It may also occur in jet fragmentation in $pp$,
$p\overline{p}$, or $e^+e^-$ collisions \cite{ellis}. 
Simply put, a Skyrmion is a configuration where the chiral field
winds nontrivially around the manifold of degenerate minima of the
effective potential.  Detailed calculations show that
the probability per unit volume to form a baryon or antibaryon
topologically is given by the following simple formula \cite{sb,early}.
\begin{equation}
{\cal P} \approx 0.08 \xi^{-3} 
\end{equation}
The interesting aspect of this mechanism is that it
produces baryons and antibaryons above and beyond those formed
by the recombination of quarks and antiquarks during a phase transition
or jet fragmentation, even when the conversion of gluons into
quark-antiquark pairs is taken into account to conserve (or create)
entropy \cite{heinz}. This mechanism is independent
of whether or not a partonic gas, or quark-gluon plasma, preceded it.

Let us now review the experimental situation.  The $\Omega$ and
$\overline{\Omega}$ were only measured in heavy ion collisions at
the CERN SPS, never at the BNL AGS.  The WA97 collaboration observed
them in Pb+Pb collisions at a beam energy of 158 A GeV within one
unit of rapidity centered around the nucleus-nucleus cm frame \cite{WA97}.
The results are quoted as $\overline{\Omega}/\Omega = 0.383\pm0.081$
and $\Omega$ + $\overline{\Omega} = 0.41\pm0.08$ per central collision.
The number of other hyperons measured are $\Xi^- = 1.5\pm0.1$,
$\overline{\Xi}^+ = 0.37\pm0.06$, $\Lambda = 13.7\pm0.9$, and
$\overline{\Lambda} = 1.8\pm0.2$.  The NA49 collaboration measured
a variety of hadrons over a much wider range of momentum space but
not, unfortunately, the $\Omega$ or $\overline{\Omega}$ \cite{NA49}.
These data were extrapolated in \cite{freeze2} to all momentum
space.  The relevant numbers are $\overline{p} = 10\pm1.7$,
$\overline{\Lambda}/\Lambda = 0.2\pm0.04$, $\Xi^- = 7.5\pm1.0$, and
$\Xi^- + \overline{\Xi}^+ = 8.2\pm1.1$.  These 4$\pi$ integrated
yields are consistent within experimental and extrapolational
uncertainties with the WA97 results.  To get an estimate of the
total number of $\overline{\Omega}$ in a central collision we
use the total multiplicity of doubly strange hyperons measured by
NA49 to convert the relative yields from WA97.
\begin{displaymath}
\left(\frac{\overline{\Omega}}{\Omega + \overline{\Omega}}   
\right)_{\rm WA97} \cdot \left(\frac{\Omega + \overline{\Omega}}
{\Xi^- + \overline{\Xi}^+}\right)_{\rm WA97} \cdot
\left(\Xi^- + \overline{\Xi}^+\right)_{\rm NA49} = 0.50
\end{displaymath}
So on average one $\overline{\Omega}$ is produced for every two
central Pb+Pb collisions.  In absolute terms this is very small
because the total net baryon number in the collision is 414.
      
The experimental data show that $\overline{p} >
\overline{\Lambda} > \overline{\Xi}^+ > \overline{\Omega}$.
We now make the extreme approximation that all
$\overline{\Omega}$ are produced as topological defects.
Since topological production creates baryons and antibaryons
in equal numbers this means that about 38\% of the $\Omega$
originate from this same mechanism.  We make the further
approximation that there is equal probability to make a
topological defect with the quantum numbers of any member
of the baryon octet or decuplet.  This is a reasonable
approximation because the topological process is not sensitive
to the mass of the defect created, as already pointed out in
\cite{heinz}.  We can phrase this another way.  Giving the $s$
quark a greater mass (110-130 MeV) than the $u$ and $d$ quarks (5-7 MeV)
suggests that the $SU(3)$ chiral field is less likely to point in the strange 
than non-strange direction.  Therefore one might conclude that $\Omega$ 
production by the defect mechanism would be suppressed relative to non-strange 
baryons.  However, this tends to be compensated by the increased production 
probability of Skyrmions when the effective potential is tilted by a nonzero 
quark mass \cite{sb}.  Given that the decuplet baryons have spin 3/2
versus spin 1/2 for the octet means that the total number of
topological defects (Skyrmions plus antiSkyrmions) created
in a typical central Pb+Pb collision is about 14.  This number
is small compared to the total number of baryons and antibaryons.
It also means that the fraction of $\Xi^-$ or $\overline{\Xi}^+$
which which were originally created via the topological mechanism
is smaller than for the triply strange hyperons.  The fraction
of singly strange hyperons is even less, and for antiprotons
the fraction is estimated to be 0.25/10 = 2.5\%.  Once created,
the nonstrange baryons for sure, and the singly and doubly strange
hyperons to a lesser degree, will still undergo some amount of
chemical equilibration.  However, as shown above and in more
detail below, the $\Omega$ and $\overline{\Omega}$ will not.  Therefore
they are a direct signal of the topological production mechanism.

To verify the indestructibility of the $\Omega$ and
$\overline{\Omega}$ following their formation, we assume that all
other hadrons have reached kinetic and chemical equilibrium,
and calculate the annihilation rates for such processes as
$\pi + \Omega \rightarrow K + \Xi$ and
$K + \Omega \rightarrow \pi + \Xi$.
Koch, M\"uller and Rafelski \cite{sreview} assembled experimental
data and assumed universal invariant matrix elements to obtain
the needed cross sections.  From figure 5.2 of their paper we
parameterized the thermally averaged product of cross section
with relative velocity for the inverse reactions
(valid for $T > 100$ as expressed in MeV).
\begin{eqnarray}
\langle \sigma(K + \Xi \rightarrow \pi + \Omega) v_{K\Xi} \rangle &=&
0.22 \,\, {\rm mb}\cdot{\rm c} \\
\langle \sigma(\pi + \Xi \rightarrow K + \Omega) v_{\pi\Xi} \rangle &=&
1.7 \left(\frac{170}{T}\right) e^{-705/T} \,\, {\rm mb}\cdot{\rm c}
\end{eqnarray}
Using these in the set of master rate equations given in their appendix A,
and using the nonrelativistic limit for the $K, \Xi, \Omega$ and the
ultrarelativistic limit for the $\pi$ (valid for $100 < T < 200$)
we obtained the following $1/e$ chemical equilibration times.
\begin{eqnarray}
\tau(\pi + \Omega \rightarrow K + \Xi) &=&
160 \left(\frac{170}{T}\right)^{3/2} e^{142.5/T} \,\, {\rm fm/c}\\
\tau(K + \Omega \rightarrow \pi + \Xi) &=&
36 \left(\frac{170}{T}\right)^2 e^{354/T} \,\, {\rm fm/c}
\end{eqnarray}
At $T = 170$ MeV, for example, these times are 370 and 290 fm/c,
respectively, far too long to annihilate any $\Omega$ or $\overline{\Omega}$.
As the temperature decreases these times grow exponentially.

The DCC domain size $\xi$ can now be estimated by equating the
total number of topological defects with the production probability
per unit volume, eq. (1), times the volume of the nuclear system
at the time of formation of topological defects.  The total number of
hadrons, both mesons and baryons, produced in central lead collisions
with rapidities within two units of midrapidity is about 2000, as
measured by NA49 \cite{mult}.  Assuming that defect production occurs
at a particle density on the order of 10 times nuclear matter density,
or 1.7 hadrons/fm$^3$, translates into a volume equivalent to a
non-Lorentz contracted lead nucleus.
This volume roughly corresponds to a collision time
in the cm frame of about 6 to 7 fm/c after first nuclear contact.
The result is $\xi$ = 2 fm.  Is this reasonable?  Essentially every
dynamical calculation of the average domain size yields a number in
the range of 1.4 to 3 fm, including thermal evolution \cite{DCC4},
quenching \cite{quench}, annealing \cite{anneal}, and bubble
nucleation \cite{bubble}.  Due to the uncertainties involved in
the present inference of $\xi$ it may be more accurate to say that
it is consistent with any value within that range.  Unfortunately
such a small value is practically undetectable in any other
observable, such as fluctuations in the ratio of neutral to
charged pions and Bose-Einstein interferometry \cite{Gavin}.

There is one other aspect to the $\Omega$ and $\overline{\Omega}$
anomaly we have not yet mentioned.  The inverse slope, $T_{eff}$,
of the transverse mass distribution of most hadrons, such as
pions, kaons, nucleons, lambdas, and deuterons all fall on a
straight line when plotted as a function of the mass $m$ of the
hadrons \cite{NA49,mT2,mT}.
\begin{equation}
T_{eff} = 180 + 105m
\end{equation}
Here $T_{eff}$ is in MeV and $m$ is in GeV.
The $\Omega$ and $\overline{\Omega}$ strongly deviate from this
systematic behavior; they have an inverse slope of $251 \pm 19$ MeV.
This is more than 5 standard deviations away from the systematics.
(There is a small deviation for the $\Xi^-$ and $\overline{\Xi}^+$
but it is much less pronounced and within 1 standard deviation of
systematics.)  The much smaller inverse slope is to be expected
if they are produced with small velocities relative to the surrounding
matter, as is the case with the formation of topological defects.   
It is difficult to make this more quantitative without doing a full
simulation.

In conclusion we have shown that the anomalies associated with
the $\Omega$ and $\overline{\Omega}$ observed in high
energy heavy ion collisions may be understood if those hyperons
are produced predominatly as topological defects in DCC.  We
infer a domain size of about 2 fm.  This is too small to affect
any of the other observables so far proposed for DCC.  This
production mechanism may arise if a quark-gluon plasma had been
formed earlier in the collision but it is not a requirement.
In fact, data obtained with lighter ions at the SPS display
a similar anomaly in the $\Omega$ and $\overline{\Omega}$
yields.  To make further theoretical progress it would seem
desirable to have a microscopic transport model which allows
for the possibility of DCC formation and the production of
topological defects.  Data from the newly commissioned
RHIC at BNL is eagerly awaited.  

\section*{Acknowledgements}

We wish to thank J.-P. Blaizot, B. M\"uller, J. Rafelski,
K. Rajagopal and J. Stachel for useful comments on the manuscript.
This work was supported by the US Department
of Energy under grant no. DE-FG02-87ER40328.


\begin{thebibliography} {99}
\bibitem{RandL} J. Rafelski and J. Letessier, nucl-th/9903018.
Unlike in most other works, such as \cite{freeze1,freeze2},
a parameter is allowed for out of equilibrium phase space occupancy
for strange quarks and another one for light quarks.  These
parameters are used whether or not the $\Omega$ and
$\overline{\Omega}$ are included in the fit.
\end{thebibliography}
\end{document}